# Modal noise mitigation for high-precision spectroscopy using a photonic reformatter

Fraser A. Pike,[1,*] Aurélien Benoît,[1] David G. MacLachlan,[1] Robert J. Harris,[2] Itandehui Gris-Sánchez,[3,4] David Lee,[5] Tim A. Birks,[3] and Robert R. Thomson[1]

[1]*SUPA, Institute of Photonics and Quantum Sciences, Heriot-Watt University, Edinburgh, EH14 4AS, UK*
[2]*Zentrum für Astronomie der Universität Heidelberg, Landessternwarte, Königstuhl 12, 69117 Heidelberg, Germany*
[3]*Centre for Photonics and Photonic Materials, Department of Physics, University of Bath, Bath, BA2 7AY, UK*
[4]*Currently with ITEAM Research Institute, Universitat Politècnica de València, Camino de Vera s/n, 46022 Valencia, Spain*
[5]*STFC UK Astronomy Technology Centre, Royal Observatory, Blackford Hill, Edinburgh, EH9 3HJ, UK*
[*]*Corresponding author: fap30@hw.ac.uk*

Recently, we demonstrated how an astrophotonic light reformatting device, based on a multicore fibre photonic lantern and a three-dimensional waveguide component, can be used to efficiently reformat the point spread function of a telescope to a diffraction-limited psuedo-slit [1]. Here, we demonstrate how such a device can also efficiently mitigate modal noise – a potential source of instability in high resolution multi-mode fibre-fed spectrographs). To investigate the modal noise performance of the photonic reformatter, we have used it to feed light into a bench-top near-infrared spectrograph ($R \approx 9{,}500$, $\lambda \approx 1550$ nm). One approach to quantifying the modal noise involved the use of broadband excitation light and a statistical analysis of how the overall measured spectrum was affected by variations in the input coupling conditions. This approach indicated that the photonic reformatter could reduce modal noise by a factor of six when compared to a multi-mode fibre with a similar number of guided modes. Another approach to quantifying the modal noise involved the use of multiple spectrally narrow lines, and an analysis of how the measured barycentres of these lines were affected by variations in the input coupling. Using this approach, the photonic reformatter was observed to suppress modal noise to the level necessary to obtain spectra with stability close to that observed when using a single mode fibre feed. These results demonstrate the potential of using photonic reformatters to enable efficient multi-mode spectrographs that operate at the diffraction limit and are free of modal noise, with potential applications including radial velocity measurements of M-dwarfs.

## 1. INTRODUCTION

Since the first discovery of a planet orbiting a main-sequence star in 1995 [2], the search for exoplanets has been a major focus of modern astronomy. Current technology can only directly image very long-period planets using the very largest telescopes, therefore detection techniques require us to look at the effect the planet has on its host star. The two most successful techniques are transit photometry [3], where the variation in the brightness of a star as the planet transits across it is measured, and the radial velocity (RV) method [4], which involves measuring the Doppler shift in the host star's spectrum due to the motion around the system barycentre. When these methods are used in combination, a great deal of information can be gathered about the exoplanet in question, including the density [5]. Typically, an exoplanet candidate is identified using a survey telescope, which relies on transit photometry. A famous example of such a telescope is the recently retired *Kepler Space Telescope*, which has allowed the identification of more than half of all confirmed exoplanets. Among others, the recently launched TESS (Transiting Exoplanet Survey Satellite) will observe the entire sky during its mission, and the PLATO mission (PLAnetary Transits and Oscillations of stars) is due for launch in 2026. They are expected to provide many new targets in the coming years. *Kepler* was designed specifically for the detection of Earth-Sun system analogues [6], and upon analysing the data, the prevalence of terrestrial planets detected was highly encouraging, although a poor detection rate of smaller radius planets was identified [7]. Following this discovery, Dressing & Charbonneau extended the *Kepler* data to smaller planets around smaller stars, and estimated that M-dwarfs hosted habitable near-Earth-sized planets at a rate of $0.15^{+0.13}_{-0.06}$ per star [8].

M-dwarfs are small cool stars (~2500 K) with peak blackbody emission in the near-infrared (NIR) range, and may have an anomalously high chance of hosting exoplanets. According to the study by Mulders, Pascucci & Apai [9], there is an inverse correlation between stellar temperature and planet occurrence rates: planets around M stars occur twice as frequently as around G stars (such as the Sun). Indeed, an M-dwarf hosts one of the largest solar systems discovered to date (apart from our own) – TRAPPIST-1 [10]. The closest exoplanet to Earth, Proxima Centauri b, also has an orbit in the habitable zone of an M-dwarf [11]. A further advantage of performing RV measurements of these cooler, lower mass stars is the increased perturbations from habitable zone planets – the habitable zone being closer and the smaller mass disparity allows much easier detection of small rocky planets which could host life [12].

To date, the most successful RV spectrograph is HARPS (High Accuracy Radial velocity Planet Searcher), which operates over visible wavelengths between 380 and 690 nm, with a resolving power of 115,000, and uses a multi-mode (MM) fibre to feed light from the telescope focal plane to the instrument, which is placed in the observatory basement where environmental conditions are strictly controlled for maximum stability [13]. Such advantages of feeding spectrographs with optical fibres are profound and well-known [14, 15]. HARPS also uses a simultaneous ultra-stable ThAr reference spectrum fed through an adjacent fibre, to allow an RV precision down to 30 cm s$^{-1}$ [16]. Unfortunately, the use of silicon-based charged-coupled-device (CCD) arrays in HARPS restricts its observations to visible wavelengths, preventing efficient RV measurements of M-dwarf stars whose blackbody emission peaks in the NIR. There is therefore a strong pull to develop high precision (~1 m s$^{-1}$) spectrographs for NIR RV measurements, but this capability still remains to be addressed. To highlight the difference in precision between wavelengths, CARMENES (Calar Alto high-Resolution search for M dwarfs with Exoearths with Near-infrared and optical Echelle Spectrographs), which saw first light in 2015, exhibits a precision of 1–2 m s$^{-1}$ in the visible, and 5–10 m s$^{-1}$ in the NIR [17]. Roy et al. [18] also state this discrepancy in precision goals, these being ~10 cm s$^{-1}$ in the visible and < 1 m s$^{-1}$ in the NIR. One example of a visible spectrograph with a precision goal of 10 cm s$^{-1}$ is ESPRESSO (Echelle SPectrograph for Rocky Exoplanets and Stable Spectroscopic Observations) [19], and an example of an NIR spectrograph with a precision goal of 1 m s$^{-1}$ is NIRPS (Near Infra-Red Planet Searcher) [20].

The use of fibres for transporting light from the telescope focal plane to the instrument is highly advantageous for environmental stability reasons, but they are not without drawbacks. For example, modal noise [21, 22] is a phenomenon that occurs where the pattern of light at the output of the fibre evolves with time, due to fluctuations in the distribution of optical energy or relative phases of the guided modes. Modal noise will arise when the stellar image at the fibre input changes (either a change in its position as the telescope slews, or from changing atmospheric conditions), or if the fibre bends (due to the telescope slewing, thermal variations or air currents). This drastically reduces the signal-to-noise [23] and severely limits the accuracy of the spectrograph. A further issue with the modal noise is that it cannot be eliminated by calibration, since the injection of the calibration source will never exactly match that of the star and thus the different coupling exacerbates the modal noise.

Modal noise is obviously minimised by using a single mode (SM) fibre to feed the spectrograph, but at the expense of telescope–fibre coupling efficiency, since atmospheric turbulence causes wavefront distortions and produces a stellar image that is not diffraction-limited. Extreme Adaptive Optics (AO) can be used to increase the coupling efficiency, and is planned for several future instruments [24]. However this is expensive, and only possible where a suitably bright natural guide star is available, limiting the number of targets [25]. In any case, the point spread function (PSF) will still not be completely diffraction-limited [26]. Once in the MM regime, modal noise actually reduces as the number of spatial modes increases due to statistical averaging. The number of modes (in a step index fibre), $N_{\text{modes}}$, is strongly dependent on the wavelength, $\lambda$, of a given fibre. This can be seen in equation (1):

$$N_{\text{modes}} = \left(\frac{\pi\, a\, NA}{\lambda}\right)^2, \quad (1)$$

where $a$ and $NA$ are the fibre core radius and numerical aperture respectively.

The simplest approach to mitigate modal noise in MM fibre-fed spectrographs is to agitate the fibre [23], which cross-couples the excited modes and averages the energy over all modes, including some that were not initially populated. This is most effectively done by hand due to the random nature of the hand movement [27], although an automated mechanical oscillator is more practical during observations. Other methods for mode-scrambling include using alternative fibre geometries e.g. octagonal or rectangular core fibres [28]. Annealed fibre has also shown effective mode-scrambling [29], since scattering centres are produced in the fibre that distribute light uniformly both radially and azimuthally. Laser speckle reducers are also available commercially and have been used to some success by Mahadevan et al. [30]. In all of these cases, however, it is important to stress that efficient mode-scrambling becomes increasingly difficult to achieve as the number of modes decreases, making modal noise a particular challenge in precision MM fibre-fed NIR spectrographs. It is for this reason that modal noise is not a significant problem for spectrographs like HARPS that operate in the visible, but was a limiting factor in GIANO at the Telescopio Nazionale Galileo (TNG) [31], which suffered from such significant modal noise that the engineers modified the entrance slit to bypass the fibre altogether [32]. This free-space approach does not take advantage of the environmental and optomechanical stability advantages offered by an optical fibre feed.

A potential alternative to standard modal-noise mitigation techniques relies on the exploitation of photonic lanterns (PLs) [33, 34], guided wave transitions that efficiently couple light from a MM port to an array of SM waveguides. PLs can be created in a variety of ways. For example, a multicore fibre (MCF) with a two-dimensional (2D) array of SM cores can be heated and tapered to form a MM port when the taper is cleaved, with the taper forming a gradual transition between the MM port and the SM cores of the MCF. Such MCF-PLs are a well-suited method to eliminating modal noise in radial velocity spectrographs [35, 36]. A second approach involves the tapering of a bundle of SM optical fibres to form a MM port in a similar manner. Regardless of the specific fabrication approach used, a spectrograph fed with MM starlight in the form of multiple SMs would be free of modal noise. To exploit the full potential of this capability, however, it is essential to correctly arrange the SMs generated by the PL at the slit of the spectrograph, such that the individual spectra from each SM do not overlap on the detector.

One approach to achieve this in MCF-PL-fed spectrographs is the TIGER approach [37], where the MCF is rotated to the correct angle, although this technique is only applicable to systems operating in the few-mode regime. Another approach proposed by Bland-Hawthorn et al. [38], known as the PIMMS (Photonic Integrated Multi-Mode Spectrograph) concept, would make use of PLs fabricated from multiple SM fibres. In this case, the individual SMs at the PL output can be arranged at will along the slit of the spectrograph.

Both the PIMMS and TIGER approaches can, in principle, enable very high-resolution MM spectrographs that exploit an Echelle grating for dispersion [39, 40]. However, for a real spectrograph system, we ideally wish to use a single MCF-PL for capturing the telescope PSF and transporting it to a spectrograph, whilst also combining this capability with the mode-reformatting flexibility offered by PLs fabricated from individual SM fibres. As we have demonstrated previously [1], one solution is the combination of an MCF-PL with a three-dimensional (3D) integrated optical waveguide mode-reformatting component fabricated using ultrafast laser inscription (ULI) – an advanced laser manufacturing technique. We have demonstrated that this combination of technologies can be seamlessly integrated together to efficiently reformat the MM PSF of the CANARY AO system operating on the William Herschel Telescope to a SM pseudo-slit, with an on-sky throughput of $53 \pm 4$ per cent in the H-band. Although the on-sky throughput results were promising, no results relating to the modal noise performance of the device were reported.

To address this, we characterised the device using a single wavelength to simulate high resolution, both experimentally and theoretically [41]. Initial findings suggested modal noise, whilst largely suppressed, was still present in the output of the device. This was in agreement with other experiments [42] and showed that to properly estimate the modal noise contribution from a photonic reformatter, a full characterisation of the device was required.

In this paper, we use a simple NIR spectrograph to investigate in detail the modal noise performance of the photonic reformatter demonstrated on sky in MacLachlan et al. [1], and compare its performance to SM and MM fibres. We call this device the "hybrid" reformatter since it integrates an MCF-PL and a ULI fabricated mode reformatting component. In Section 2 we will outline the experimental design of the spectrograph and the experimental techniques used to investigate the modal noise performance. In Section 3 we describe the data processing methods and the results obtained, which demonstrate that near-SM performance can be obtained using the hybrid reformatter. In Section 4 we link the precision of our device to potential scientific applications.

## 2. EXPERIMENTAL SPECTROGRAPH DESIGN AND DATA CAPTURE METHODS

We have designed an inexpensive bench-top spectrograph constructed from catalogue components, which enables observation and quantification of modal noise using different optical fibre feeds. For our purposes, light was fed into the spectrograph using three devices; the hybrid reformatter (HR), an SM fibre patch cord (SMF28e, FC/PC connectors at both ends) with a mode field diameter (MFD) of $10.4 \pm 0.5$ μm at 1550 nm and mode NA of 0.14, and a step index MM fibre with 50 μm core diameter and NA of 0.22. All devices had a length of approximately 2 m. A schematic of the spectrograph design, images of the end-facets of the three devices, and their respective typical output light patterns are presented in Fig. 1. A schematic of the complete HR is presented in Fig. 2. It is crucial to highlight that the number of guided modes supported in this 50 μm MM fibre at 1550 nm is around 124, similar to the 92 modes of the lantern, which has a core size of 43 μm, to give the fairest comparison possible.

For all characterisation experiments, the light source used was an SM fibre-coupled broadband amplified spontaneous emission (ASE) source (Thorlabs FL70002-C4) centred on 1560 nm, close to the centre of an M-dwarf emission spectrum. A bandpass filter (BPF) (Thorlabs FB1550-12) with a full width at half maximum (FWHM) of 12 nm was also used to ensure that the spectrum was within one free spectral range of the Echelle Grating, eliminating the need for cross-dispersion. Neutral density (ND) filters were placed in the input beam path when necessary to avoid saturating the camera. Lens L1 is a 15.58 mm focal length fibre collimation package (NA 0.16) and lens L2 is a 10 mm focal length and 8 mm diameter achromatic doublet, producing an image of the fibre mode which has an MFD of $6.7 \pm 0.5$ μm. This image is initially aligned with the input to the device under test (DUT) by maximising the throughput. Lens L3 has 25.4 mm focal length, and lens L4 has 500 mm focal length for a magnification of $\approx 20$. This magnification has been selected to allow the image of the pseudo-slit created by the HR to fill the majority of the detector height. The Echelle Grating (EG) was sourced from Thorlabs (GE2550-0363) with 63º blaze angle and 31.6 lines $mm^{-1}$; this is used due to the high efficiency in higher orders, which greatly increases the dispersion and is essential to achieving a high resolution. M1 is a 2" square silver mirror used to fold the beam path and overcome restrictions from bulky optic mounts. This allows the grating to be used close to its design angle (the Littrow configuration). Finally, the output raw images are recorded by a Hamamatsu C10633-23 InGaAs camera, based on a detector cell of 256×320 pixels with a pixel size of 30×30 μm. The resolving power $R$ of the spectrograph is $\lambda/\Delta\lambda$, where $\Delta\lambda$ is the FWHM of the line function. The resolving powers we calculate for each DUT are as follows: $R \approx 9{,}500$ for SM fibre, $R \approx 3{,}500$ for MM fibre, and $R \approx 7{,}000$ for the HR.

To mimic the effect of the atmosphere and telescope slewing as stars are tracked across the sky, the coupling into the fibre was adjusted in a semi-random meander across the facet while maintaining a throughput greater than 80 per cent of the maximum at optimal coupling. At each fibre position, 250 frames were captured at 15 ms exposure time with the camera. These frames were then added together to produce a single detector image for that fibre position, simulating a longer and thus more relevant exposure time. The input coupling is modified successively to obtain 60 of these images, which were then processed according to Appendix A. In the following section we discuss how the data were analysed using two different experimental protocols to quantify the modal noise when characterising each DUT. We also investigate the effect of shaking the DUT on the modal noise performance. To do so, a mechanical scrambling system was constructed by fixing the centres of 2 independently controlled loudspeakers to loops of the DUT, which is represented by S in Fig. 1(a).

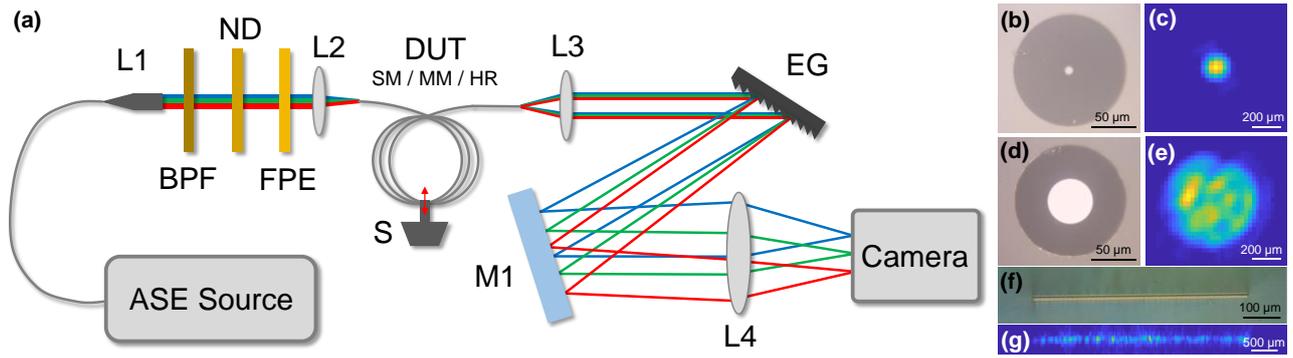

**Fig. 1.** (a) Schematic of spectrograph layout including fibre-coupled ASE source, L1 fibre collimation package, BPF bandpass filter, ND filter, FPE Fabry-Pérot etalon (if required), L2 coupling into the device under test (DUT), mechanical shaking system S, collimation by L3, the dispersion by an Echelle Grating (EG), folded by mirror M1 and focussed by L4 onto an InGaAs Camera; (b) image of the 8.2 μm core SM fibre; (c) typical output light pattern of the SM fibre; (d) image of the 50 μm core MM fibre; (e) typical output light pattern of the MM fibre; (f) image of the 570 μm long pseudo-slit; (g) typical output light pattern of the single-mode pseudo-slit. Note that the image of the pseudo-slit is not to scale with the fibres, and the output light patterns are imaged through lenses which magnify the image.

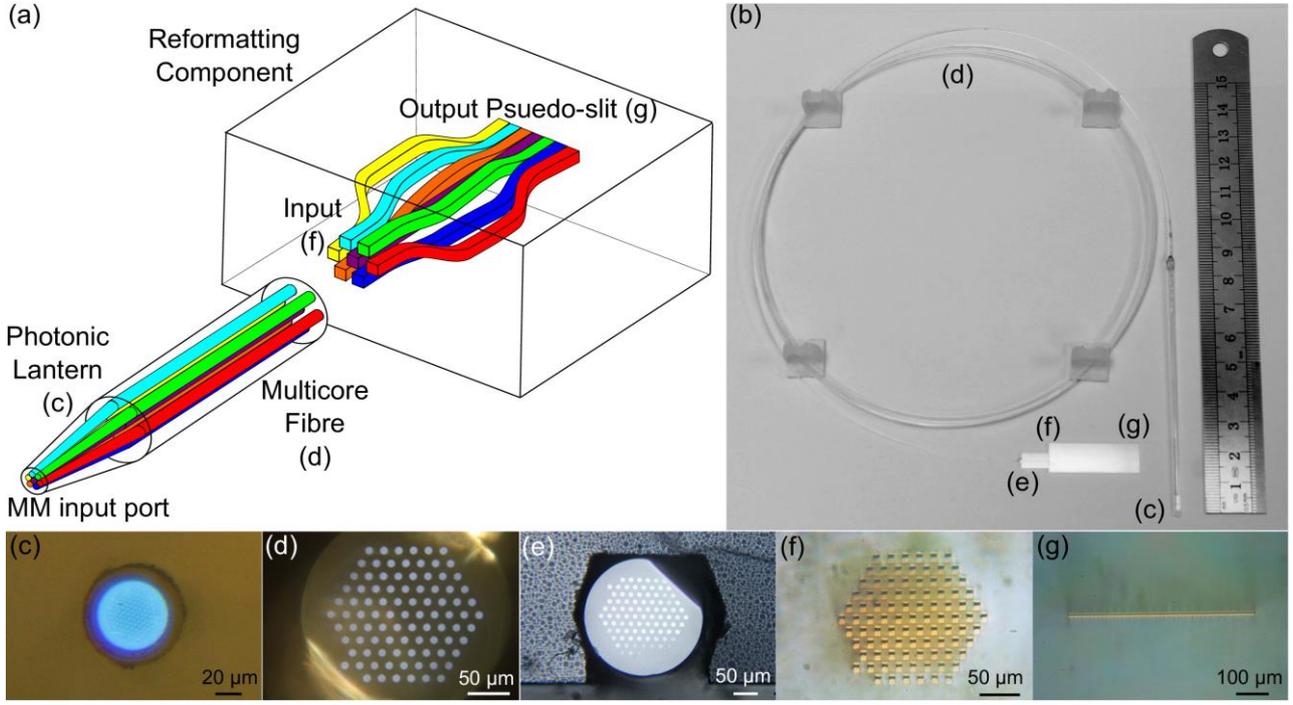

**Fig. 2.** Key: (a) Schematic diagram of the complete device with the different colours showing the paths of the SMs; (b) Photograph of the device with 15 cm ruler for scale; (c) MM input facet of the PL; (d) MCF of 92 cores in a hexagonal array; (e) MCF placed in a V-groove; (f) input facet of the ULI reformatter chip; (g) SM pseudo-slit output. From MacLachlan et al. [1].

## 3. DATA PROCESSING TECHNIQUES AND RESULTS

### A. Characterising modal noise from a broadband measured spectrum

In our characterisation system, modal noise manifests itself through changes in the measured spectrum as the input coupling is varied. The strength of the modal noise can therefore be determined by quantifying how the spectrum varies across the full data set. To do so, we compare the spectrum obtained from each of the 60 images to the mean spectrum across all images.

A higher degree of modal noise will result in larger differences, which can then be quantified statistically. More specifically, 60 spectra were obtained by summing each processed image along the spatial axis (as described in Appendix A). Figs. 3(a-d) illustrate examples of raw data (unprocessed images) acquired using a single input coupling position for each of the 3 DUTs, including examples where the MCF of the HR is either static or shaken during the exposure, and when shaken we define this as HRS. Only the data obtained using the MM fibre with shaking (MMS) is shown because the visual difference in the images taken using with and without shaking was low by eye. Figs. 3(e–h) presents spectra obtained from the raw data sets after processing, with the black line representing the average spectrum obtained across the 60 different input coupling positions, and the red line representing the spectrum out of the 60 that was most different from the average. The peak of the spectrum appears at slightly different positions on the camera from one DUT to another due to small differences in the physical positioning of the DUT relative to the spectrograph. There is no reason to expect these differences will affect the analysis.

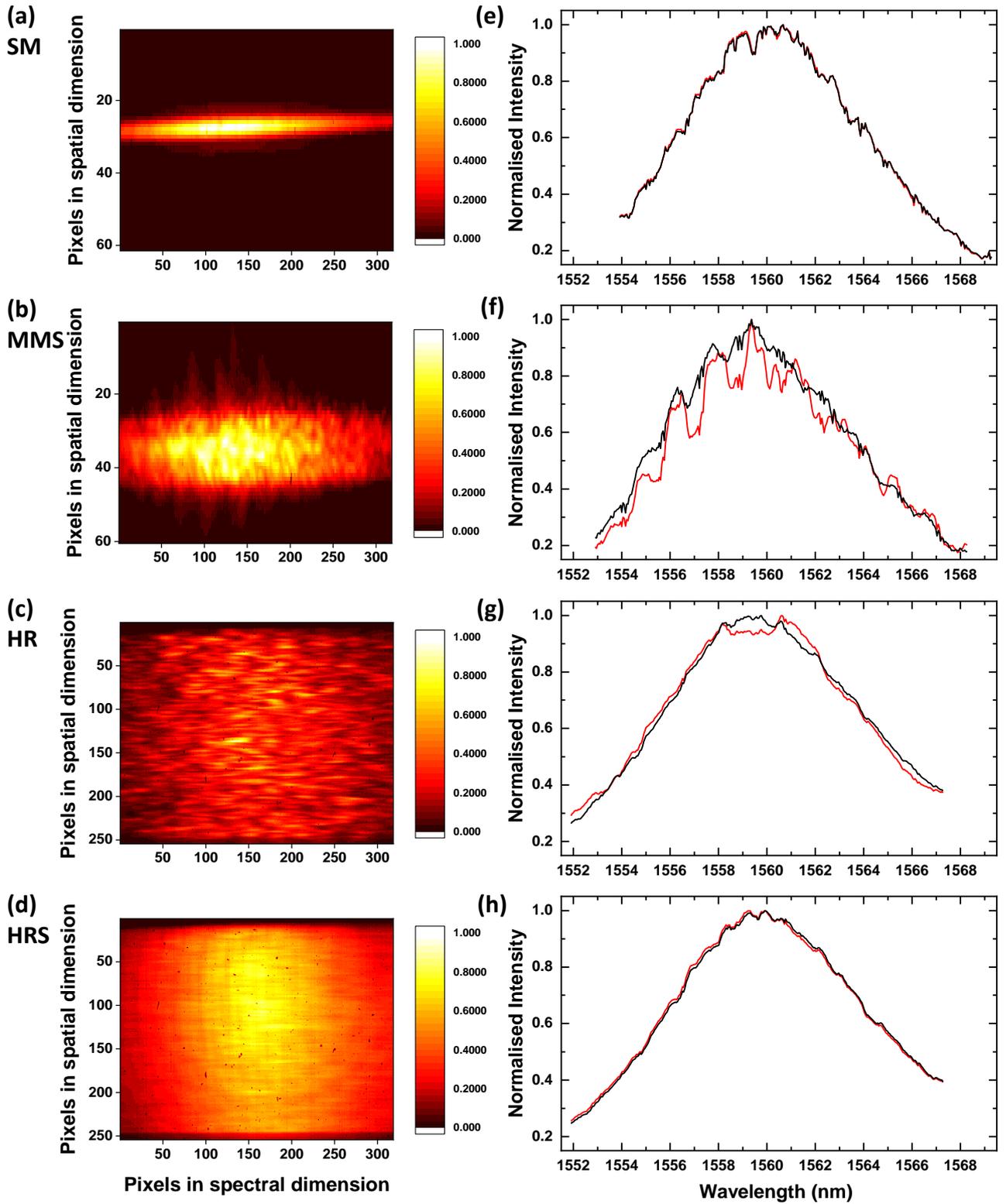

**Fig. 3.** Examples of normalised coloured raw images for: (a) SM, (b) MMS, (c) HR and (d) HRS; Mean spectrum (black line) and the most different spectrum (red line) for (e) SM, (f) MMS, (g) HR and (h) HRS. Each raw image corresponds to the most different spectrum. Some empty space on the detector with the SM and MM fibres has been omitted. Note that the normalisation step shown here is shown for the benefit of the reader only, and does not form part of the data analysis.

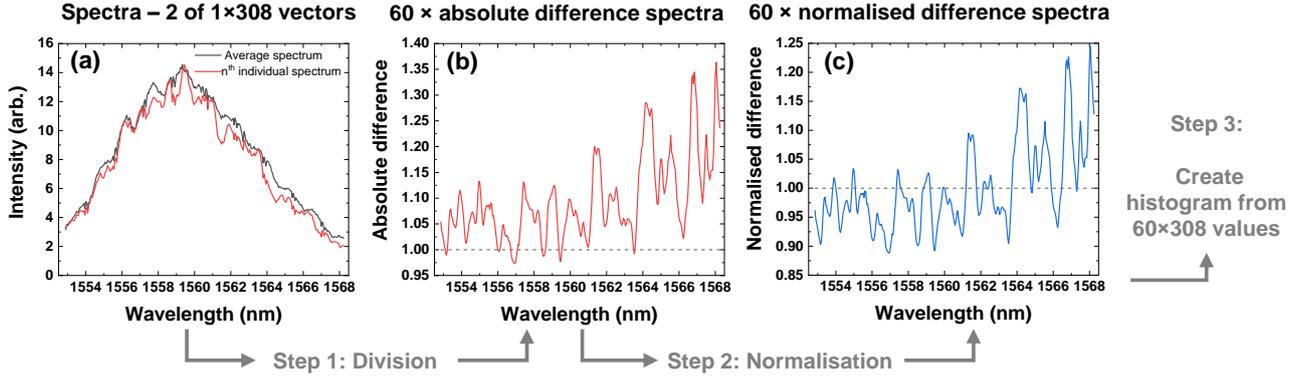

**Fig. 4.** The three steps to obtain the histogram in Fig. 5 are illustrated using MMS data: (a) shows how the $n^{th}$ individual spectrum compares to the mean; (b) shows the absolute difference spectrum obtained by dividing these; (c) shows the normalised difference spectrum. A histogram is then formed from the full data set of 60×308 elements.

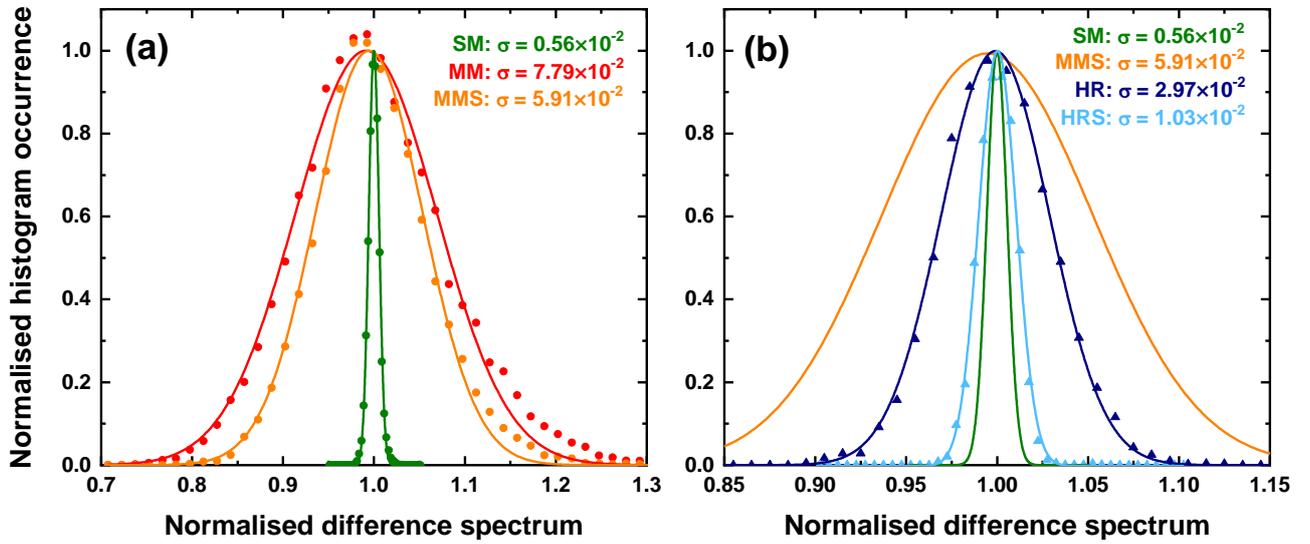

**Fig. 5.** The data represents the histogram of the difference between spectra at different coupling positions, with a wider distribution indicating larger differences in intensity for each wavelength unit. The solid lines are a Gaussian fit to the data. (a) Normalised occurrence of difference spectrum value under changing coupling in MM (red circles), MMS (orange circles) and SM (green squares) fibres; (b) Normalised occurrence of difference spectrum value under changing coupling in the HR (dark blue triangles) and in the HRS (light blue triangles) superimposed onto SM (green curve) and MMS (orange curve).

**Table. 1.** Parameters of the fits to difference spectrum data in Fig. 5. We present $\sigma$ with a factor of $10^{-2}$ removed for ease of comparison.

| Device | Standard Deviation $\sigma$ (×$10^{-2}$) | $R^2$ goodness-of-fit |
|---|---|---|
| SM | 0.56 | 0.998 |
| MM | 7.8 | 0.989 |
| MMS | 5.9 | 0.995 |
| HR | 3.0 | 0.997 |
| HRS | 1.03 | 0.996 |

The difference between the data and spectra obtained using the SM fibre (Figs. 3(a&e)) and the MMS data (Figs. 3(b&f)) is immediately apparent. Here we see the characteristically smooth spectrum obtained using the SM fibre, and the highly variable spectrum obtained using the MM fibre, the shape of which will vary with input coupling. The difference between the spectra obtained using the HR and HRS devices are also observable, and we highlight the impact of the scrambling process in Figs. 3(c&d). The HRS image appears very similar to the SM image spectrally, but with a large difference in the height of the area filled on the detector due to the length of the pseudo-slit. Supplementary videos will be made available with the journal submission showing the evolution of the 60 images for each DUT.

The data processing steps followed to quantify the modal noise using the acquired data are depicted in Fig. 4 to aid the reader. First, by elementwise-dividing the average spectrum by the $n^{th}$ (out of 60) measured spectrum (Fig. 4(a)), we obtain a 308 value vector that represents the absolute deviations between the $n^{th}$ spectrum and average spectrum (Fig. 4(b)). This vector is then normalised such that the mean value is 1, which accounts for any variation in the absolute power of the $n^{th}$ spectrum (Fig. 4(c)). The deviation from unity in this vector is related to the modal noise. Once this process is applied to the full data set, a histogram of the 18480 values, sorted into 40 bins, can be plotted to represent the spectral differences across all the 60 spectra due to modal noise. As seen in Fig. 5, the histograms (data points) that are generated by this data processing are well approximated by a Gaussian distribution, the best fit of which is also presented. We have normalised the histograms so that the Gaussian fit has a peak value of 1 to allow a straightforward comparison. The strength of the modal noise can be represented by the standard deviation of the fitted Gaussian, $\sigma$, and these can be found in Table 1, along with the goodness-of-fit.

It is apparent that the modal noise present in the spectra measured using the HRS is greatly reduced compared to when using either the HR or MMS devices, and is approaching the performance using the SM fibre. It should be highlighted that the width of the histogram obtained using the SM fibre represents the experimental limits of our characterisation system. We see that the data obtained using the HR exhibits a factor of 2.6 reduction in modal noise compared to the data obtained with the MM device, and a factor 2.0 reduction in modal noise compared to the data obtained using the MMS device. Shaking the MCF of the HR reduces the modal noise by a further factor of 2.9. The data obtained using MMS exhibits only a factor 1.3 reduction in modal noise compared to the data obtained using the MM fibre. This brings the total modal noise mitigation to a factor 5.7 between the HRS and the MMS – using the same scrambling system, with a similar number of guided modes. These results clearly demonstrate the superior scrambling ability of the HR device in comparison to the MM fibre, and demonstrates our first method to quantify how a photonic reformatter such as the HR can efficiently mitigate modal noise.

### B. Characterising modal noise from the barycentre precision of spectral peaks

The results outlined in Section 3.1 provide a straightforward route to quantify modal noise and modal noise mitigation, but do not provide an immediate quantification of how modal noise affects the precision of a spectrograph. To address this, we have investigated a second method to quantify modal noise. This method uses the same characterisation system shown in Fig. 1, but with a Fabry-Pérot etalon placed in the input beam path between L1 and L2, converting the smoothly varying broadband light source into a series of discrete spectral peaks spaced by a regular frequency interval – the etalon free spectral range. As outlined previously, modal noise generates variations in the acquired spectra as the input coupling is varied. Here, we use the measured spectral stability of the etalon peaks under different input coupling conditions as a proxy for the strength of modal noise using different DUTs. The etalon we chose was sourced from LightMachinery and was made from solid fused silica with a thickness of 0.821 mm and surface reflectivities of ~0.885 and ~0.873 respectively. These parameters produce an etalon with a finesse of 23, generating spectral peaks with a width of approximately 40 pm spaced by ≈ 1 nm at 1550 nm. This etalon was chosen since the spectral peaks are sufficiently spaced such that they can still be resolved by the spectrograph when fed using light via the MM fibre.

In Figs. 6(a–d) we present the raw images of data acquired using three DUTs (again with both unshaken and shaken conditions for the MCF of the PL). In Figs. 6(e–h) we present how the acquired spectrum varies with input coupling, again with the solid black line indicating the average spectrum for 60 measurements, and a sample of 5 of these spectra at different coupling positions represented by the coloured dashed lines. Again, since there is a lack of visual difference between the MM fibre with and without shaking, only the MMS measurements are presented in Fig. 6. We again used the process outlined in Appendix A to correct for deviations in the straightness of the pseudo-slit and the angle between the pseudo-slit and the pixel axes of the camera.

For each of the spectral peaks generated by the etalon a Gaussian fit was made to determine the central wavelength (barycentre). The fit was typically over 10 pixels which gave a high confidence in the precision. The variation in the acquired spectra is due to modal noise while varying the input coupling, which in turn results in variations in the measured barycentres of each peak. Thus, the standard deviation of the 60 measured barycentres for each peak is our second measure of modal noise. This data is plotted in Figs. 7(a–b), where the $x$ axis position of each data point represents the average measured barycentre for an etalon peak, and the $y$ axis position of each data point is the standard deviation of the 60 measured barycentres for that spectral peak.

It is also useful to plot these barycentres as a ratio of the width of the peaks themselves, as this can link the precision to the resolution, plotted in Fig. 7(c). We calculate the peak width from the FWHM of the Gaussian fit. The large uncertainty for the MM/MMS measurements is due to the non-Gaussian shape of the peaks providing an imprecise fit. In Table 2 we present the mean values (and associated uncertainty from the standard deviation of those 16 values) of the barycentre stability over each of the 16 peaks, calculated from the standard deviation (SD) of 60 barycentres as seen in Fig. 7. It is apparent that the photonic approach using HRS offers a significant improvement over the MM fibre and a performance close to that measured using the SM fibre, and also that the mode scrambling system is highly effective – reducing the modal noise by a factor of 5 compared to when using the HR without shaking. The variation of the barycentre across the pseudo-slit with the HRS is 0.56 per cent of the peak width. The effect of shaking the MM fibre is again observed to be minimal compared to when shaking the HR. The mean SD as a percentage of the peak width for the SM fibre is equal to the HRS, showing their equivalent performance in this spectrograph.

### C. Correcting for variations in the laboratory temperature

Our experiments were performed in a basic lab without the 0.01 K temperature control and vacuum chambers used in a state-of-the-art spectrograph. Temperature effects can therefore introduce instabilities in the spectrum that are not due to modal noise. For example, the true spectral positions of the etalon peaks may drift by 10 pm K$^{-1}$ [43]. With the aim of accounting for the impact of laboratory thermal fluctuation on our barycentre method of quantifying modal noise we have also conducted the following additional analysis.

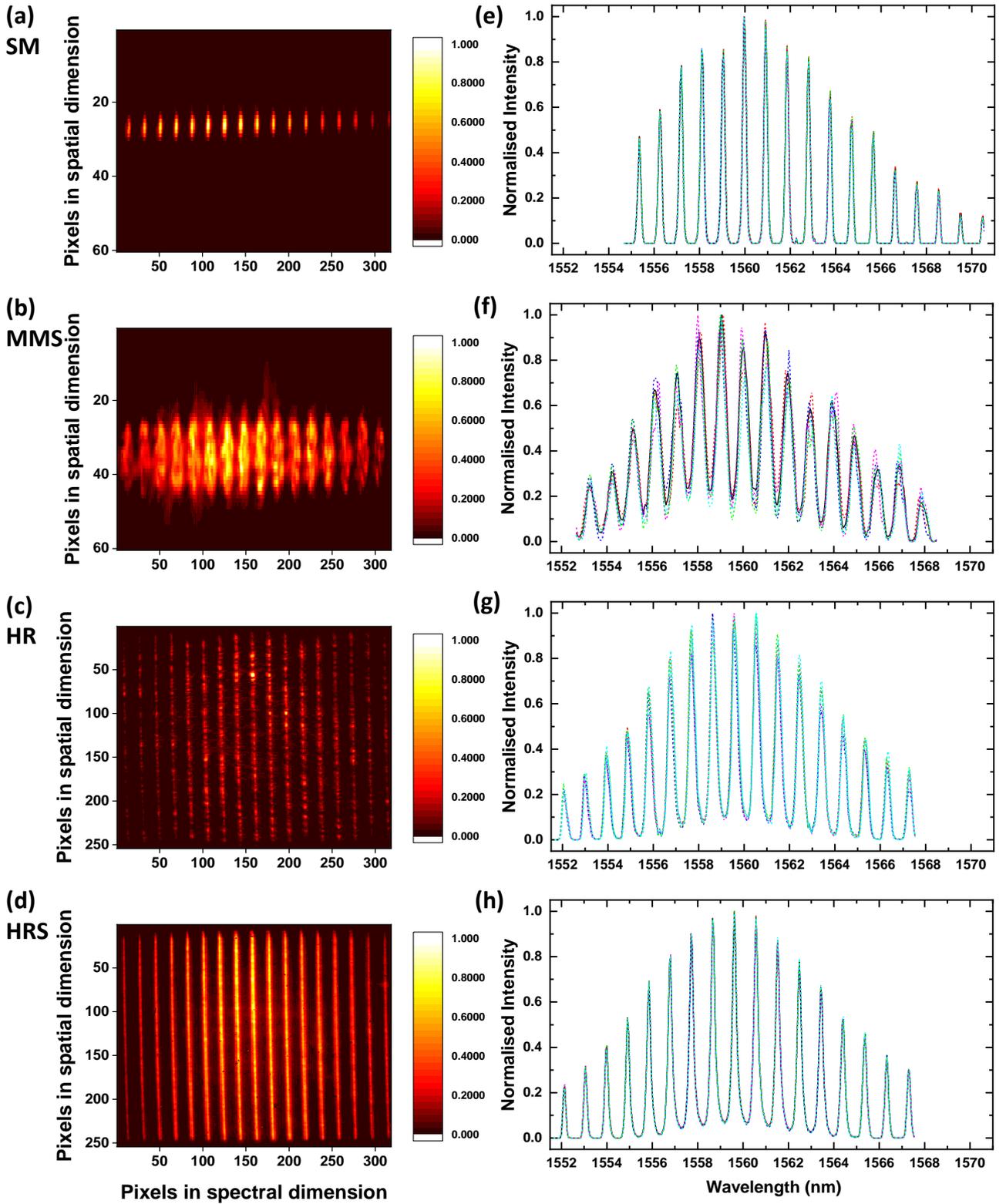

**Fig. 6.** Examples of normalised coloured raw images with the etalon for: (a) SM, (b) MMS, (c) HR and (d) HRS; Examples of 5 randomly chosen spectra (dashed line) and the mean spectrum (solid line) for (e) SM, (f) MMS, (g) HR and (h) HRS. Again, we present raw images of the most different spectra, and some empty space on the detector with the SM and MM fibres has been omitted.

It is logical to assume that laboratory temperature drifts will have a very similar effect on the spectral peaks across the measurement, since the wavelength span of the measurement is very small. Therefore, by examining the manner in which the measured spectrum shifts for each of the 60 measurements compared to the mean spectrum measured across all 60, it should be possible to substantially account and correct for laboratory temperature variations.

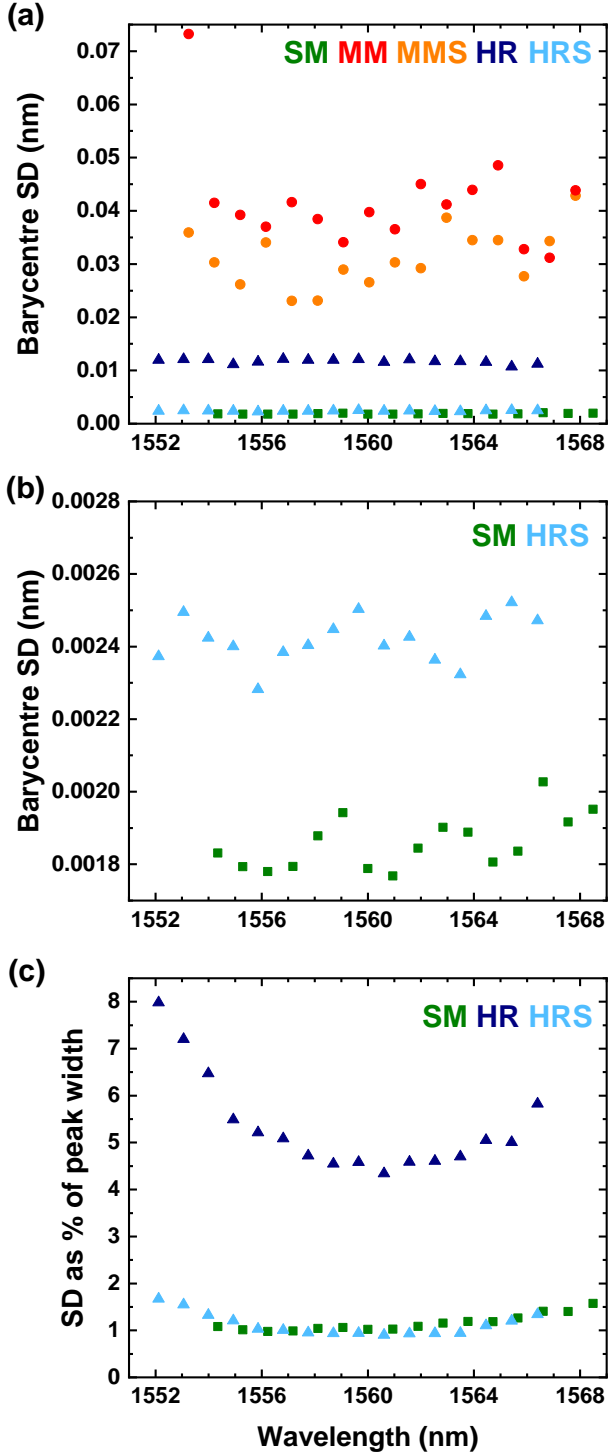

**Fig. 7.** (a) Standard deviation of 60 calculated barycentres for each spectral peak at the given average wavelength, where green squares correspond to SM, red circles to MM, orange circles to MMS, dark blue triangles to HR, light blue triangles to HRS; (b) Close-up of SM and HRS in (a); (c) Standard deviation of 60 calculated barycentres, plotted as a percentage of the respective peak's width, for each spectral peak at the given average wavelength. The spectra of each DUT were sampled at different positions which causes the slight offset on the graph.

**Table 2.** Average values of the barycentre precision presented in Fig. 7. We present the SD with a factor of $10^{-3}$ removed for ease of comparison.

| Device | Mean SD (×$10^{-3}$ nm) | Mean SD as % of peak width |
|---|---|---|
| SM | 1.86 ± 0.07 | 1.2 ± 0.2 |
| MM | 42 ± 10 | 8 ± 2 |
| MMS | 31 ± 6 | 6 ± 2 |
| HR | 11.7 ± 0.4 | 5.3 ± 1.0 |
| HRS | 2.42 ± 0.07 | 1.1 ± 0.2 |

For a given spectrum, we use the average deviation of 15 etalon peaks from their mean positions (across the 60 measurements) as a proxy to represent how much the spectrograph has drifted from its mean position due to thermal effects – we call this shift the "temperature proxy". The temperature proxy was observed to gradually vary over the 60 measurements with a full range of ~8 pm. This would, for example, correspond to an etalon temperature range of slightly less than one degree. We then spectrally shift each of the 60 spectra by the temperature proxy, such that the mean position of 15 peaks is the same across all 60 measurements. This will primarily compensate for variations in the laboratory temperature, leaving mainly the instability in the peak positions due to modal noise. A Fourier transform of the difference spectrum between the mean spectrum in Fig. 3(h) and any one of the 60 contributing spectra indicates that the modal noise in the HRS occurs with a period that is ~25 per cent shorter than the period of the etalon peaks. This means that every third etalon peak samples the modal noise with the same phase, meaning that the effect of the modal noise on the mean barycentre shift is negligible when considering every group of three adjacent etalon peaks. We therefore calculate the temperature proxy using the 15 etalon peaks which lie closest to the centre of the wavelength span (an integer of 3), rather than the 16 available, so that we do not subtract contributions to the mean barycentre shift that are due to modal noise.

The standard deviation of the position of each of the etalon peaks relative to their respective mean positions can then be recalculated, and plotted to generate "laboratory temperature corrected" versions of Figs. 7(b&c), presented as Figs. 8(a&b). In Table 3 we present the mean values of the data shown in Fig. 8. When compared to the data presented in Table 2, it is clear that accounting for the effect of laboratory temperature increases the stability of the etalon spectral peaks by a factor of ~6 for both the SM and HRS DUTs. It is also interesting to note the shape of the curves shown in Fig. 8, where the etalon peaks are observed to be most stable in the middle of the spectral range. This, we believe, is due to the spectral peaks being physically wider on the detector array in the centre of the spectral range, becoming progressively narrower to either side (the difference between the widest and narrowest peaks are factors of 1.5 and 1.9 for SM and HRS respectively). This may be due to field curvature resulting from lens L4 which prohibits all wavelengths from simultaneously being in focus on the flat detector, and Zemax simulations support this belief. There is no reason to suggest that it will not be possible, with a more carefully engineered spectrograph, to achieve the stability observed with the etalon peaks in the centre of the spectral range.

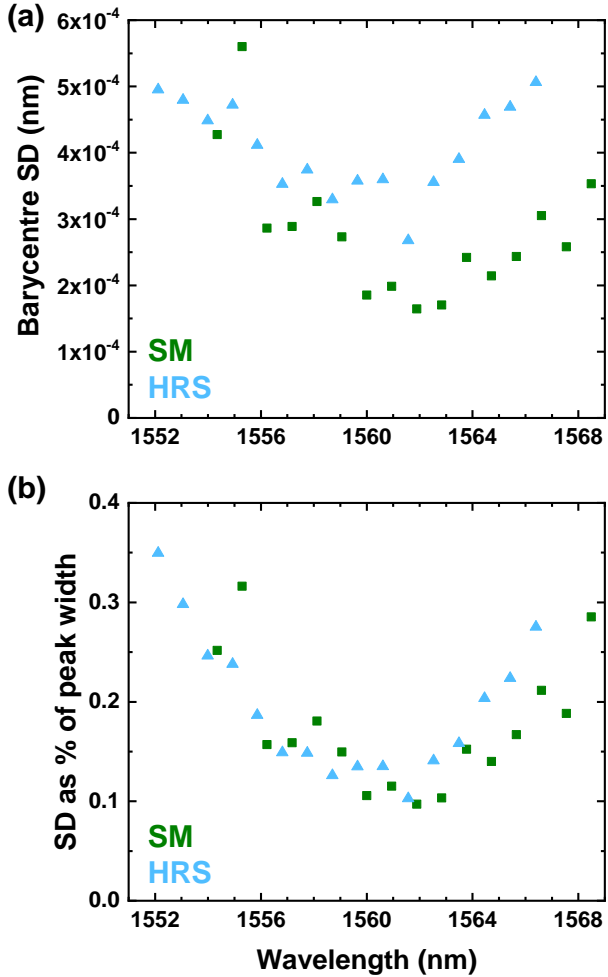

**Fig. 8.** (a) Standard deviation of 60 calculated barycentres when laboratory temperature correction is applied to the data, for each spectral peak at the given average wavelength, where green squares correspond to SM and light blue triangles to HRS; (b) Standard deviation of 60 calculated barycentres when laboratory temperature correction is applied to the data, plotted as a percentage of the respective peak's width, for each spectral peak at the given average wavelength. The spectra of each DUT were sampled at different positions which causes the slight offset on the graph.

**Table 3.** Average values of the barycentre precision presented in Fig. 8. We present the mean SD with a factor of $10^{-3}$ removed for ease of comparison.

| Device | Mean SD (×$10^{-3}$ nm) | Mean SD as % of peak width |
|---|---|---|
| SM | 0.28 ± 0.10 | 0.17 ± 0.06 |
| HRS | 0.41 ± 0.07 | 0.19 ± 0.07 |

Fig. 8(b) indicates that the barycentres of the etalon peaks at around 1559 nm are stable to a thousandth of the width of the peak for both the SM and HRS DUTs. Based on a barycentre precision of 8 ± 2 per cent of the peak width for the MM fibre DUT, the HRS was observed to result in a factor of ~100 improvement in the barycentre stability. Anagnos et al. [44] have used beam propagation simulations to model the propagation of light through a photonic reformatting component similar to the HR device investigated here, and concluded that they should increase the barycentre precision by a factor of 1000 compared to a 50 μm core MM fibre. It is important to note, however, that since the data we have obtained using the SM DUT represents the modal noise measurement limit of our characterisation system and methods. Fig. 8(b) therefore merely demonstrates that HRS exhibits a level of modal noise that is not detectable using our experimental system and the methods we have described. With an improved experimental system and optimised experimental protocol, it is logical to expect that the graphs presented in Fig. 8 will both reduce in magnitude further and eventually separate, with the data for the SM DUT dropping further.

## 4. COMMENTS ON ACHIEVABLE RADIAL VELOCITY PRECISION

When using the HRS and our benchtop spectrograph, we are able to achieve a barycentre stability of $0.41 \times 10^{-3}$ nm after accounting for the effect of laboratory temperature variations. This would infer that a single spectroscopic line could be measured to an accuracy of ~80 m s$^{-1}$, assuming all other sources of noise are negligible. We therefore conclude that if our spectrograph (or similar) were placed in an environmentally controlled container, it would already operate with a precision close to that required for scientific applications e.g. for detecting hot Jupiters such as WASP-19b [45], which orbits a star of 12$^{th}$ magnitude and requires a high-throughput spectrograph such as one enabled by this device. If the HRS was used to feed light into a higher resolution spectrograph (such as ≈ 120,000 offered by the current state-of-the-art) using a camera with smaller pixel size (15×15 μm is the current state of the art) we might expect a 0.1 per cent barycentre stability relative to the physical peak width to result in a single line radial velocity precision around 1 m s$^{-1}$, again assuming all other sources of noise are negligible. This is easily low enough to detect a terrestrial exoplanet in the habitable zone around an M-dwarf. We also note the fact that real RV measurements are almost always made by cross-correlating full spectra consisting of many spectral lines, and so it is reasonable to conclude that the achievable precision when limited by modal noise could be significantly higher.

## 5. CONCLUSIONS

We have developed a bench top near infrared spectrograph to characterise the modal noise performance of a photonic reformatter called the hybrid reformatter (HR) which reformats a telescope point spread function to a diffraction-limited pseudo-slit. We used the spectrograph to compare the modal noise performance of the HR to that exhibited by two reference devices: a single mode fibre and a multi-mode fibre which supported a similar number of guided modes. We also investigated the effect of mechanical shaking on the modal noise.

We used two methods to quantify the strength of the modal noise. In the first we used a spectrally smooth broadband source and a statistical analysis to quantify how the entire acquired spectrum changed as a result of different input coupling conditions to simulate the effect of telescope slewing and tracking during an exposure. Using this method, we observed that the modal noise performance of the HR when shaken was a factor of $\approx 6$ better than that observed using the MM fibre when shaken, but a factor of $\approx 2$ worse than when using the SM fibre.

In the second, we used a broadband source consisting of multiple spectrally narrow peaks to quantify how the barycentres of the peaks shift as a result of different input coupling conditions. In this case, we observed that the modal noise performance of the HR when shaken was identical to that of the SM fibre, but we again highlight that this merely indicates that the HR when shaken exhibits a level of modal noise that is not detectable using our experimental system and the barycentre method we have described.

Finally, looking forward to science applications, we have considered the relevance of our modal noise characterisation tests in the context of NIR radial velocity measurements, concluding that HR devices could offer a powerful route to combine high throughput efficiencies enabled by multi-mode operation with high precision spectroscopy through strong modal noise mitigation.

## Acknowledgements

This work was funded by the UK Science and Technology Facilities Council (STFC) – STFC grant no. ST/N000625/1, and by the European Union's Horizon 2020 research and innovation program under grant agreement No. 730890 (OPTICON – Optical Infrared Coordination Network for Astronomy). F. A. Pike acknowledges support via an EPSRC iCASE studentship part funded by Renishaw.

The raw data will be made available on Heriot-Watt University's Pure data repository.

**Appendix A. Process to create spectra from raw data**

After gathering the 60 images, some initial data processing was performed. The 5 hot pixels identified on the camera were all set to zero, and the outermost rows and columns of the camera images were removed from the data as they did not register any light, which results in a matrix of 254×318 elements. Additionally, there is a small but non-trivial angle between the peaks formed by the etalon when the psuedo-slit is imaged on the detector, and the spatial axes of the camera; this is most easily noticeable in Fig. 6(d). The process to "straighten" the image is explained here with the aid of a schematic in Fig. A1. A calibration file was determined from the data gathered using the HRS, since this is more reliable than the non-shaken conditions. Starting with one image of 250 rows (we eliminate the top 4 empty rows to achieve a multiple of 10) (Step 1), we obtain a partial spectrum for each block of 10 rows in an image (summed along the axis orthogonal to the dispersion), which we can compare to the 250-row spectrum (Step 2). We perform a cross-correlation of the two spectra, and the lag corresponding to the peak of the cross-correlation is where they show most similarity (Step 3). We did this using summed sections of 10 rows because a cross-correlation using individual rows was prone to error. Applied to the full data set, this lag, corresponding to the $m^{th}$ block of rows in the $n^{th}$ image, was set as element $(m,n)$ in a 25×60 matrix (Step 4). This calibration matrix was then averaged down the 60 columns and rounded to an integer to determine how much each block of rows should be shifted (Step 5). The range of the shift required was ~7 pixels over the full length of the pseudo-slit. The calibration file was applied consistently to all 60 images in both the shaken and unshaken data sets. The resolution increased by 50 per cent – becoming more comparable to the resolution of the spectrograph when fed with SM fibre. The "straightened" images were then summed down the spatial axis to create 60 spectra as before. This process was not used to correct the data from the SM fibre as the small angle and minimal number of rows spatially occupied by the image made the process needless. In addition to this reason being applicable to the MM fibre (albeit to a lesser extent), the modal noise present meant this method of cross-correlation was unreliable and actually decreased the resolution. In this case it would also not be reasonable to use the same calibration file as for the pseudo-slit because the spectrograph needed some small physical adjustments to accommodate the differently shaped outputs. The calibration file was retrospectively applied to the broadband images of the pseudo-slit without the etalon to ensure the statistical analysis we perform in section 3.1 used the spectrograph at the same resolution. The wrapping of rows when shifted caused some errors on the edges of the images, so the outer 5 columns on each side were truncated to ensure no unnecessary errors in the spectra were introduced by the data processing. For each DUT, to determine the relationship between pixel number and wavelength a tuneable laser (Anritsu MG9638A) was scanned across the detector in 1 nm increments. A straight line fit then gave the conversion factor (since the small wavelength range does not introduce non-linear dispersion).

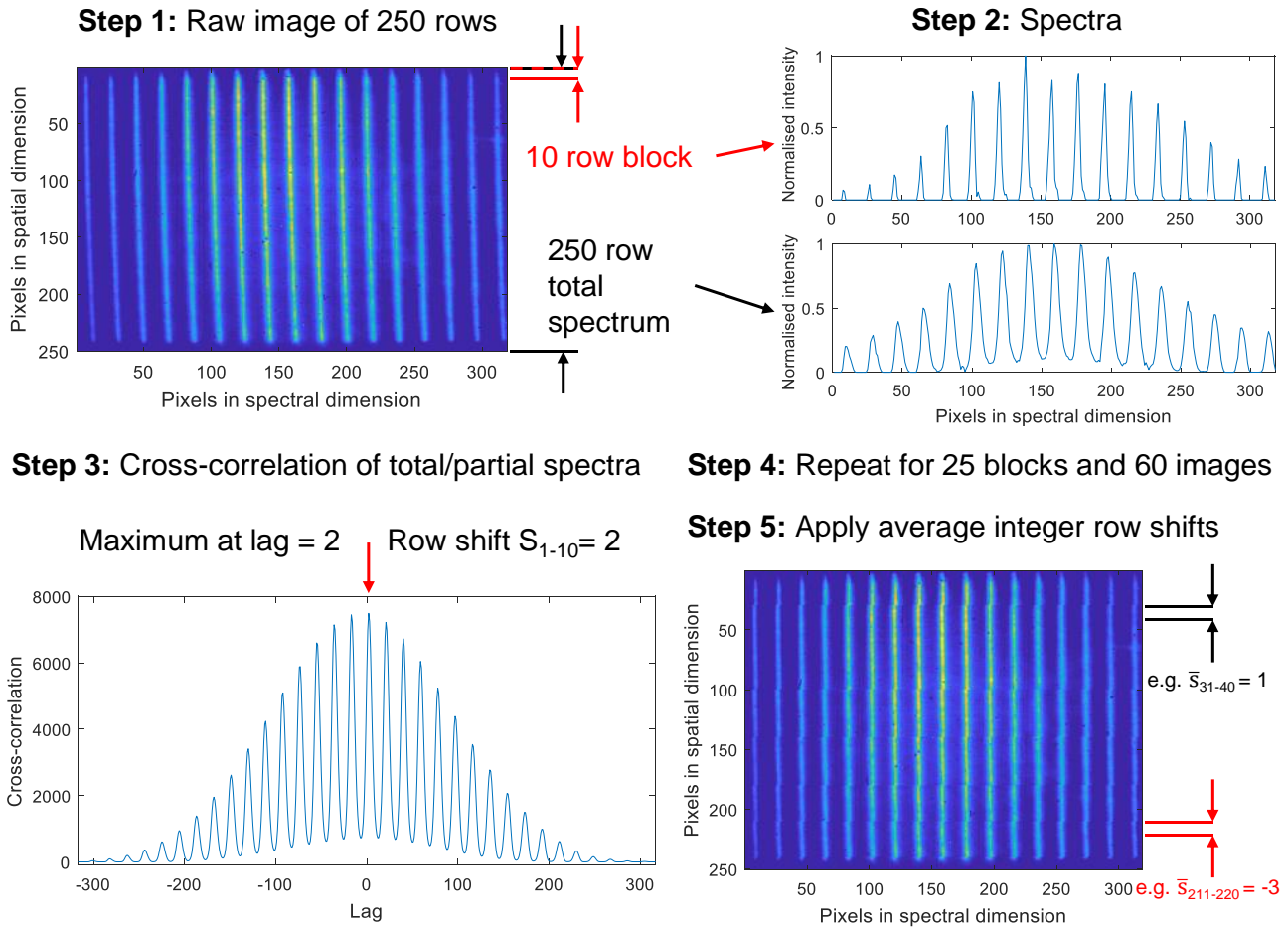

**Fig. A1.** The five steps to straighten the image of the pseudo-slit on the detector are presented. Step 1 is to select a raw image of 250 rows. Step 2 is to generate the total spectrum and a partial spectrum from a 10-row block. Step 3 is to perform a cross-correlation between these spectra. The lag corresponding to the maximum value is the row shift for that block and particular image. Step 4 is to repeat this for 25 blocks and 60 images to generate a matrix. The average row shift for each block is calculated and rounded to the nearest integer. Step 5 is to apply the row shifts for each block consistently for all images.